\documentclass[preprint,showpacs,nofootinbib]{revtex4}
\usepackage{amsmath}
\usepackage{graphicx}
\usepackage{physics}
\usepackage{hyperref}
\usepackage{color}

\newcommand{\diff}{\mathop{}\!\mathrm{d}}

\begin{document}

\title{Chapman-Enskog calculation of the shear viscosity of
  quark-gluon plasma including all $2\leftrightarrow 2$
  scatterings at finite temperature} 
\author{Okey Ohanaka}
\affiliation{Department of Physics, East Carolina University,
  Greenville, NC 27858, USA} 
\author{Zi-Wei Lin}
 \email{linz@ecu.edu}
\affiliation{Department of Physics, East Carolina University,
  Greenville, NC 27858, USA} 

\date{\today}

\begin{abstract}
We use the Chapman-Enskog method to investigate the shear viscosity of
the quark-gluon plasma with a focus on its relation to parton cross
sections. We use the recently obtained analytical expression for the
shear viscosity $\eta$ of a massless quark-gluon gas at chemical
equilibrium with Boltzmann statistics and all $2\leftrightarrow 2$
scatterings with arbitrary cross sections. Here we apply this general
expression to cross sections at finite temperature that are based on 
perturbative-QCD and screened with scaled thermal masses
$\sqrt{\kappa}\,m_D$ and $\sqrt{\kappa}\,m_F$. 
We find that the Chapman-Enskog results on $\eta \, g^4/T^3$  versus
$m_D/T$ at $\kappa=1$ are qualitatively similar to but higher than the
corresponding leading-order results from the AMY framework.  We then
find that using $\kappa=0.4$ allows the Chapman-Enskog results to
match well the corresponding AMY results as it includes the effect of
using thermal masses (instead of self-energies) to screen the cross
sections. In addition, we show that the shear viscosity-to-entropy
density ratio $\eta/s$ is very sensitive to the choice of momentum
scale $Q$ in the strong coupling, where the choice of $Q=3T$ leads to
$\eta/s \sim 0.15$  for $N_f=0$ or 3 at the QCD phase transition
temperature $T_c$. These results lay the foundation for mapping parton
cross sections to given shear viscosity in parton transport models and
QCD effective kinetic theory.
\end{abstract}

\maketitle

\section{Introduction}\label{Intro}

The quark-gluon plasma (QGP) is a state of matter created when the
energy density is sufficient for quarks and gluons to become
deconfined and then reach or come close to equilibrium. 
For the early stage of high energy heavy ion collisions or certain
small collision systems where the lifetime of the parton phase is not
long enough for sufficient equilibration, an off-equilibrium parton
matter is expected to exist. To study the evolution of the dense
matter, especially the non-equilibrium dynamics of the quark-gluon
system, parton transport models have been developed, such as Zhang's
parton cascade (ZPC)~\cite{Zhang:1997ej}, a multi-phase transport  
(AMPT)~\cite{Lin:2004en}, Boltzmann approach to multi-parton
scatterings (BAMPS)~\cite{Xu:2007ns}, parton-hadron-string dynamics
(PHSD)~\cite{Cassing:2009vt}, quasiparticle model
(QPM)~\cite{Plumari:2011mk}, and invariant or covariant parton 
cascades~\cite{Kurkela:2022qhn,Nara:2023vrq}. 
Models of QCD effective kinetic theory~\cite{Kurkela:2018vqr} have
also been developed. They are important tools for studies of high
energy nuclear collisions including the reliable extraction of the QGP 
properties such as the shear
viscosity~\cite{csernai2006,Bernhard:2019bmu}. For
example, it has been  realized that parton transport and kinetic
theory generate significant anisotropic flows through the parton escape
mechanism~\cite{He:2015hfa,Lin:2015ucn,Kurkela:2018ygx} even when the
collision number per parton is on the order of one or lower.

Since shear viscosity is a key property of the quark-gluon matter
including the QGP, it will be useful to have an analytical expression
of the shear viscosity in terms of parton scattering cross
sections. This will allow us to easily calculate the shear viscosity
of a parton transport or kinetic theory and to extract the shear
viscosity from the comparisons with experimental data. The shear
viscosity in QCD has been calculated in many works before
~\cite{Arnold:2000dr,Policastro:2001yc,Arnold:2003zc,Kovtun:2004de,Meyer:2007ic,Ghiglieri:2018dib}, 
including the well known results at high temperature from the AMY
framework~\cite{Arnold:2000dr, Arnold:2003zc}, where the
perturbative-QCD (pQCD) scattering matrix  elements are being
regulated with parton self-energies and both $2\leftrightarrow 2$ and  
$1\leftrightarrow 2$  processes are included. 
Since parton scattering cross sections have large uncertainties,
especially for physically relevant temperatures below several $T_c$, a
general analytical expression of the shear viscosity for arbitrary
differential cross sections would be very useful, and
such an analytical expression for all $2\leftrightarrow 2$ elastic and
inelastic scatterings has been derived in our recent work using the
Chapman-Enskog (CE) method~\cite{Ohanaka:2026hjx}. In this study, we
apply this general expression to the specific pQCD-based
$2\leftrightarrow 2$ cross sections to study the corresponding shear
viscosity of the QGP and its relation to the parton scattering cross
sections. 

\section{The Shear Viscosity in terms of All $2\leftrightarrow 2$ 
  Cross Sections}
\label{etaCE}

In a recent study~\cite{Ohanaka:2026hjx}, we have obtained the
general expression of the shear viscosity of a multi-species massless
quark-gluon gas in chemical equilibrium as 
\begin{equation}
\eta = 160\, T ~ \frac{x_g^2 \left[C_{qq}^{00} + C_{q\bar{q}}^{00} +
    2(N_f-1)C_{qq^{\prime}}^{00}\right] - 4 N_f x_g x_q\, C_{gq}^{00} +2 N_f
  x_q^2 \,C_{gg}^{00}} {C_{gg}^{00}\left[C_{qq}^{00} +
    C_{q\bar{q}}^{00} + 2(N_f-1)C_{qq^{\prime}}^{00}\right] - 2N_f
  \left(C_{gq}^{00}\right)^2}.
\label{etaNf}
\end{equation}
In the above, $T$ is the temperature, and $x_i$ is the number fraction
of parton species $i$ (from 1 to $1+2N_f$ for a system of massless 
gluons and $N_f$ flavors of quarks). For chemical equilibrium, we have
written $x_{q_j} \equiv x_q$ and $x_{\bar {q_j}} \equiv x_{\bar q}$ for
$j \in [1,N_f]$, with $x_q=x_{\bar q}=3 (1-\delta_{0N_f})/(8+6N_f)$
and $x_g=8/(8+6N_f)$. The matrix elements are written in the 
following forms: 
\begin{align}
C_{ii}^{00}= x_i^2\,\Tilde{c}_{0}[i] + x_i \sum_{j \neq i}
x_j \, \Tilde{c}_{1}[ij] + \Tilde{\omega}[ii], \; \;
C_{ij}^{00} &= x_i x_j \, \Tilde{c}_{2}[ij] + \Tilde{\omega}[ij] 
\; \; ({\rm for}~i\neq j).
\end{align}

For completeness, we list below the matrix elements for massless
partons. With $v \equiv \sqrt{s}/T$ where the Mandelstam variable $s$
is the squared two-parton center-of-mass energy, the elastic terms are
given as follows:
\begin{equation}
\Tilde{c}_{0}[i] = \frac{1}{384} \int_0^{\infty}
\diff v\, v^6 \left [ (3v^2+4)K_3(v) - 6v\,
      K_2(v) \right ] \sigma_{\rm tr}^{ii\rightarrow ii}(v), 
\label{c0g}
\end{equation}
\begin{multline}
    \Tilde{c}_{1}[ij] = - \frac{1}{192} \int_0^{\infty}
\diff v\, v^2 \, \int_{-v^2 T^2}^{0} \diff t\, \tfrac{t}{T^2} 
    \bigg\{\left[3v^2\left(\tfrac{t}{T^2} + v^2\right) +
      4\left(\tfrac{t}{T^2} + 6v^2\right)\right] K_3(v) \\ 
    - 2v\left(\tfrac{3t}{T^2} - 2v^2\right) K_2(v)\bigg\}
    \frac{\diff\sigma}{\diff t}^{ij\rightarrow ij}, 
\end{multline}
\begin{multline}
    \Tilde{c}_{2}[ij] = -\frac{1}{192} \int_0^{\infty}
\diff v\, v^2 \, \int_{-v^2 T^2}^{0} \diff t\, \tfrac{t}{T^2} 
    \bigg\{\left[3v^2\left(\tfrac{t}{T^2} + v^2\right) +
      4\left(\tfrac{t}{T^2} - 4v^2\right)\right] K_3(v) \\ 
    - 2v\left(\tfrac{3t}{T^2} + 8v^2\right) K_2(v)\bigg\}
    \frac{\diff\sigma}{\diff t}^{ij\rightarrow ij}. 
\end{multline}
The inelastic terms are given as follows:  
\begin{equation}
    \Tilde{\omega}[gg] = \frac{x_g^2 N_f}{192} \int_0^{\infty} \diff
    v\, v^6 \left[(v^2+28)K_3(v) - 2v\, K_2(v)\right]
    \sigma^{gg\rightarrow q\bar{q}}(v),
\label{wgg}
\end{equation}
\begin{multline}
\Tilde{\omega}[gq] = \frac{x_g^2}{768} \int_0^{\infty} \diff v\, v^6
\left[(3v^2+4)K_3(v) - 6v\, K_2(v)\right] \sigma_{\rm tr}^{gg\rightarrow
  q\bar{q}}(v) \\ 
- \frac{x_g^2}{384} \int_0^{\infty} \diff v\, v^6
\left [ (v^2+28)K_3(v) - 2v\, K_2(v) \right ] 
\sigma^{gg\rightarrow  q\bar{q}}(v), 
\end{multline}
\begin{equation}
    \Tilde{\omega}[qq] = \frac{x_q
      x_{\bar{q}}}{384} \int_0^{\infty} \diff v\, v^6
    \left[(v^2+48)K_3(v) + 8v\, K_2(v)\right] 
    \left[\sigma^{q\bar{q}\rightarrow gg}(v) +
      (N_f-1)\sigma^{q\bar{q}\rightarrow q^{\prime} \bar{q}^{\prime}}(v)\right], 
\end{equation}
\begin{equation}
    \Tilde{\omega}[q\bar{q}] = \frac{x_q x_{\bar{q}}}{384}
    \int_0^{\infty} \diff v\, v^6 \left[(v^2+8)K_3(v) - 12 v\,
      K_2(v)\right] \left[\sigma^{q\bar{q}\rightarrow gg}(v) +
      (N_f-1)\sigma^{q\bar{q}\rightarrow q^{\prime} \bar{q}^{\prime}}(v)\right], 
\end{equation}
\begin{multline}
    \Tilde{\omega}[qq^{\prime}] = \frac{x_q x_{\bar{q}}}{768}
    \int_0^{\infty} \diff v\, v^6 \left[(3v^2+4)K_3(v) - 6v\,
      K_2(v)\right] \sigma_{\rm tr}^{q\bar{q}\rightarrow q^{\prime}
      \bar{q}^{\prime}}(v) \\  
    - \frac{x_q x_{\bar{q}}}{384} \int_0^{\infty} \diff v\, v^6
    \left[(v^2+28)K_3(v) - 2v\, K_2(v)\right]
    \sigma^{q\bar{q} \rightarrow q^{\prime} \bar{q}^{\prime}}(v). 
\label{wqqprime}
\end{multline}
Note that the total cross section and the total transport cross
section of an individual scattering channel $ij \rightarrow kl$ in
the above relations are given respectively by  
\begin{equation}
\sigma^{ij \rightarrow kl}(v)=
\frac{1}{1+\delta_{kl}} \int_{-s}^0 {\frac {d\sigma}{dt}}^{ij \rightarrow
    kl} dt, ~~ \sigma_{\rm tr}^{ij \rightarrow
  kl}(v)=\frac{1}{1+\delta_{kl}} \int_{-s}^0 {\frac {d\sigma}{dt}}^{ij
  \rightarrow kl} \sin^2\!\theta_{\rm cm} \, dt, 
\label{dsigmadt}
\end{equation}
where $\theta_{\rm cm}$ is the scattering angle in the
two-parton center-of-mass frame. In chemical equilibrium, we have
$\Tilde{\omega}[gq] =\Tilde{\omega}[qg] = \Tilde{\omega}[g\bar{q}]
=\Tilde{\omega}[\bar{q}g]$, 
$\Tilde{\omega}[qq] = \Tilde{\omega}[\bar{q}\bar{q}]$,
$\Tilde{\omega}[q\bar{q}]= \Tilde{\omega}[\bar{q} q]$,  
and $C_{ij}^{00}=C_{ji}^{00}$.

\section{Parton $2\leftrightarrow 2$ Cross Sections at Finite
  Temperature}
\label{crossSections}

Parton scattering cross sections based on leading-order matrix
elements in vacuum often diverge, while the cross sections in a
finite temperature medium can be regulated or screened by the parton
self-energies~\cite{Arnold:2000dr, Arnold:2003zc}. 
In this study, we use a simpler scheme by screening the cross sections
with parton thermal masses. For diagrams with a gluon exchange, we
use the Debye mass $m_D = gT \sqrt{1 +N_f/6}$ while for a fermion
exchange we use the Fermion mass $m_F =gT/\sqrt{6}$, where $g^2
=4\pi \alpha_s$. We take the following leading-order strong coupling: 
\begin{equation}\label{alphas}
\alpha_s(Q^2) = \frac{4\pi}{\left(11 - \frac{2}{3}N_f\right)
  \ln(Q^2/\Lambda^2)} \; , 
\end{equation}
where $Q$ is the momentum transfer and $N_f$ is the number of active
quark flavors. For temperatures of interests for current nuclear collisions
(below $\sim 4T_c$ with $T_c \simeq 156$ MeV~\cite{HotQCD:2018pds}),
we are mostly concerned with three quark flavors ($u,d,s$). 
In this study, we take $\Lambda = 250$ MeV, where this $\Lambda$ value
is chosen so that Eq.\eqref{alphas} for $N_f=3$ fits well the
experimental extractions of
$\alpha_s(Q^2)$~\cite{ParticleDataGroup:2022pth}.  
For the Boltzmann distribution of massless partons, we have
$\langle s \rangle = 18 T^2$ and $\langle -t \rangle = 9 T^2$ for the 
collision of two partons, with $s$ and $t$ being the Mandelstam
variables. Therefore, we take $Q=\sqrt{\langle   -t \rangle}=3T$ by
default in this study, noting that there are other choices of
$Q$~\cite{Blaizot:2001tn,Laine:2005ai,Ghiglieri:2018dib}. 

For brevity, we write the elastic cross section
$\sigma^{ij\rightarrow   ij}$ as $\sigma_{ij}$; we also write the
inelastic cross sections as $\sigma_{q\bar q}^{gg} \equiv \sigma(q\bar 
q \rightarrow gg)$, $\sigma_{q\bar q}^{q^{\prime} \bar q^{\prime}}
\equiv \sigma(q\bar q \rightarrow q^{\prime} \bar   q^{\prime})$, and 
$\sigma_{gg}^{q\bar q} \equiv \sigma(gg \rightarrow q\bar
q)$, where $\sigma_{gg}^{q\bar q} =9\, \sigma_{q\bar q}^{gg}/32$. We
write the matrix elements with similar abbreviated notations. 
Since the cross sections in a finite temperature medium are
screened, we start from the following matrix
elements (summed over the spins and colors of all four
particles)~\cite{Arnold:2000dr,Arnold:2003zc} by screening the 
propagators with scaled thermal masses:
\begin{equation}
\abs{\mathcal{M}_{gg}}_0^2 =
    1152g^4 \left [ 3-\frac{s\,u}{(t-\mu_D^2)^2} - \frac{s\,t}{(u-\mu_D^2)^2} -
  \frac{t\,u}{s^2} \right ],
\label{mgg0}
\end{equation}
\begin{equation}
\abs{\mathcal{M}_{gq}}_0^2 =96g^4 \, \frac{(s^2 +
    u^2)}{(t-\mu_D^2)^2}  - \frac{128g^4}{3}
\,\left(\frac{s}{u-\mu_F^2} + \frac{u}{s}\right),
\label{mgq0}
\end{equation}
\begin{equation}
\abs{\mathcal{M}_{qq}}_0^2 = 16g^4 \left [ \frac{s^2 +
  u^2}{(t-\mu_D^2)^2} + \frac{s^2 + t^2}{(u-\mu_D^2)^2}\right ] -
  \frac{32g^4}{3} \frac{s^2}{t\,u},
\label{mqq0}
\end{equation}
\begin{equation}    
\abs{\mathcal{M}_{q \bar{q}}}_0^2 = 16g^4\left [ \frac{s^2
    +u^2}{(t-\mu_D^2)^2} +  \frac{t^2 +u^2}{s^2}\right ]
-\frac{32g^4}{3}\frac{u^2}{s\,t},
\label{mqqbar0E}
\end{equation}
\begin{equation}
\abs{\mathcal{M}_{q\bar{q}}^{gg}}_0^2 =
    \frac{128g^4}{3} \,\left(\frac{u}{t-\mu_F^2} + \frac{t}{u-\mu_F^2}\right) -
   96g^4 \, \left ( \frac{t^2 + u^2}{s^2} \right ),
\label{mqqbar0A}
\end{equation}
\begin{equation}
\abs{\mathcal{M}_{q\bar q}^{q^{\prime} \bar q^{\prime}}}_0^2 =16g^4
\left ( \frac{t^2 + u^2}{s^2} \right ),
\label{mqqbar0B}
\end{equation}
\begin{equation}
\abs{\mathcal{M}_{q q^{\prime} }}_0^2 =16g^4 \, \frac{(s^2 +
  u^2)}{(t-\mu_D^2)^2}. 
\label{mqqp0}
\end{equation}
We call above the original matrix elements and have thus labeled them
with the subscript $0$. Note that the screening in the above is
performed by replacing the relevant $t$ (or $u$) variable with
$t-\mu^2$ (or $u-\mu^2$), where 
\begin{equation}
\mu_D=\sqrt {\kappa} \, m_D =\sqrt {\kappa}\, gT \sqrt{1
  +\frac{N_f}{6}}, ~\mu_F=\sqrt {\kappa} \, \frac{gT}{\sqrt{6}}, 
\label{kappa}
\end{equation}
and $\kappa$ can be called the screening coefficient. 
A common screening method is to subtract $m_D^2$ or $m_F^2$ from the
relevant Mandelstam variables, equivalent to $\kappa=1$. Because the
AMY framework~\cite{Arnold:2000dr, 
  Arnold:2003zc} uses parton self-energies for the screening while we
use the simpler method of screening with thermal masses (which is
known to affect the shear viscosity results~\cite{Arnold:2003zc}), we
introduce the multiplicative constant $\kappa$ to take care of this
difference. With details provided in Sec.\ref{Results}, we take
$\kappa=0.4$ in this study unless specified otherwise. 

We find that two of the above original matrix elements,
$\abs{\mathcal{M}_{q\bar{q}}^{gg}}^2$ and $\abs{\mathcal{M}_{qq}}^2$,
have  negative values in certain $s-t$ region~\cite{Ohanaka:thesis},
which lead to negative $\sigma$ and/or $\sigma_{\rm tr}$ at small
$s$. Let us take $\abs{\mathcal{M}_{q\bar{q}}^{gg}}^2$ as an example: 
as $s/\mu_F^2 \rightarrow 0$, its $s-$channel term remains finite
while the $t-$ and $u-$channel terms approach zero, which leads to
negative values for the matrix element.  Note that this issue of
negative values is unrelated to the $\kappa$ value. 
To make these matrix elements always non-negative, we thus add
``minimal'' screenings to the last term of these
two matrix elements~\cite{Ohanaka:thesis} as shown below:
\begin{equation}
\abs{\mathcal{M}_{qq}}^2 = 16g^4 \left [ \frac{s^2 +
  u^2}{(t-\mu_D^2)^2} + \frac{s^2 + t^2}{(u-\mu_D^2)^2}\right ] -
  \frac{32g^4}{3} \frac{s^2}{(t-\mu_D^2) (u-\mu_D^2)},
\label{mqq}
\end{equation}
\begin{equation}
\abs{\mathcal{M}_{q\bar{q}}^{gg}}^2 =
    \frac{128g^4}{3} \,\left(\frac{u}{t-\mu_F^2} + \frac{t}{u-\mu_F^2}\right) -
   96g^4 \, \frac{(t^2 + u^2)}{s(s+\mu_D^2)}.
\label{mqqbarA}
\end{equation}

We also find that four of the other five original matrix elements (except for
$\abs{\mathcal{M}_{q q^{\prime} }}_0^2$)  lead to cross sections that
either diverge at all $s$ or diverge as $s \rightarrow 0$. For
example, the  $1/(s\,t)$ term in $\abs{\mathcal{M}_{q \bar
 q}}_0^2$ of Eq.\eqref{mqqbar0E} leads to a $\int_{-s}^0dt/t$ term in
$\sigma_{q \bar q}$ that makes it divergent at all $s$, while
$\abs{\mathcal{M}_{gg}}_0^2$ of Eq.\eqref{mgg0} leads to $\sigma_{gg}
\propto 1/s$ at small $s$ that diverges as $s \rightarrow 0$. Since a
main purpose of this study is to allow finite-temperature parton cross
sections to be mapped to given shear viscosity in parton transport
models such as ZPC~\cite{Zhang:1997ej} and AMPT~\cite{Lin:2004en}, 
finite cross sections are needed. Therefore, we add extra screenings
to the divergent terms in the matrix elements. Specifically, we add
screening to the $1/t$ term in $\abs{\mathcal{M}_{q \bar
    q}}_0^2$ to get the following modified matrix element that gives a  
finite $\sigma_{q\bar q}$ at finite $s$,  
\begin{equation}    
\abs{\mathcal{M}_{q \bar{q}}}^2 = 16g^4\left [ \frac{s^2
    +u^2}{(t-\mu_D^2)^2} +  \frac{t^2 +u^2}{s (s+\mu_D^2)}\right ]
-\frac{32g^4}{3}\frac{u^2}{s (t-\mu_D^2)}.  
\label{mqqb}
\end{equation}
where we also replaced $1/s^2$ with $1/[s(s+\mu_D^2)]$ to get finite
$\sigma_{q\bar q}$ at $s=0$. In addition, to make the other cross
sections finite at $s=0$,  we add screenings to mostly the $1/s$ terms   
by replacing $1/s$ with $1/(s+\mu_D^2)$ or $1/(s+\mu_F^2)$
(depending on the exchange particle type).
For example, for the original $\abs{\mathcal{M}_{gg}}_0^2$ we replace
the constant term $3$ with $3s/(s+\mu_D^2)$ and replace the term
$tu/s^2$ with $tu/[s(s+\mu_D^2)]$ to get the following final matrix
element (i.e., with fully screenings):  
\begin{equation}
\abs{\mathcal{M}_{gg}}^2 =1152g^4 \left [
  \frac{3s}{s+\mu_D^2}-\frac{s\,     u}{(t-\mu_D^2)^2} - \frac{s\,
    t}{(u-\mu_D^2)^2} -\frac{t u}{s (s+\mu_D^2)} \right ]. 
\label{mgg} 
\end{equation}
Similarly, we get the other matrix elements with fully screenings as
follows:
\begin{equation}
\abs{\mathcal{M}_{gq}}^2 =96g^4 \, \frac{(s^2 +
    u^2)}{(t-\mu_D^2)^2}  - \frac{128g^4}{3}
\,\left(\frac{s}{u-\mu_F^2} + \frac{u}{s+\mu_F^2}\right),
\label{mgq} 
\end{equation}
\begin{equation}
\abs{\mathcal{M}_{q\bar q}^{q^{\prime} \bar q^{\prime}}}^2 =16g^4\,
\frac{(t^2 + u^2)}{s(s+\mu_D^2)}.
\label{mqqbarB} 
\end{equation}
Note that the original $\abs{\mathcal{M}_{q q^{\prime}
  }}_0^2$ of Eq.\eqref{mqqp0} already gives a finite and non-negative
$\sigma_{q q^{\prime}}$ at any $s$ value and thus does not need any 
modification.

For the scattering cross sections of massless partons, we have 
\begin{equation}
    \frac{\diff\sigma}{\diff t} = \frac {\overline{
        \abs{\mathcal{M}}^2}}{16 \pi s^2}, 
\end{equation}
where $\overline{\abs{\mathcal{M}}^2}$ represents the squared matrix
element after averaging over the spins and colors of the two incoming 
partons. Using the final matrix elements listed in
Eqs.\eqref{mqqp0}-\eqref{mqqbarB}, we obtain the following total
cross section of each $2 \rightarrow 2$ scattering channel according  
to Eq.\eqref{dsigmadt}:
\begin{equation}
\sigma_{gg} = \frac{3g^4}{128\pi s} \Bigg[\frac{17s}{s+\mu_D^2}+
\frac{12s}{\mu_D^2} - 12 \ln  \!\left (1+\frac {s}{\mu_D^2} \right )
\Bigg] , 
\label{siggg}
\end{equation}
\begin{equation}
\sigma_{gq} = \frac{g^4}{144\pi s} \Bigg[ 11+\frac{18s}{\mu_D^2}
+\frac{9 \mu_D^2}{s+\mu_D^2} -\frac{2 \mu_F^2 }{s+\mu_F^2} -\! 18 \!
\left (   1+\frac{\mu_D^2}{s} \right ) \ln \! \left (1+\frac
  {s}{\mu_D^2} \right ) + 4 \ln \! \left (1+\frac {s}{\mu_F^2} \right
) \Bigg] ,  
\end{equation}
\begin{equation}
\sigma_{qq}=\frac{g^4}{108\pi s} \Bigg[
3+\frac{6s}{\mu_D^2}+\frac{3\mu_D^2}{s+\mu_D^2} 
- 2 \left
  (4+\frac {3\mu_D^2}{s} -\frac {2\mu_D^2}{s+2\mu_D^2} \right )  \ln \! \left
  (1+\frac   {s}{\mu_D^2} \right ) \Bigg], 
\end{equation}
\begin{equation}
\sigma_{q\bar{q}} = \frac{g^4}{27\pi (s+\mu_D^2)} \Bigg[
2+\frac{3s}{2\mu_D^2}+\frac{\mu_D^2}{4s} -\frac{\mu_D^4}{2s^2}  - \left
  (1+\frac {3\mu_D^2}{2s} -\frac {\mu_D^6}{2s^3} \right )  \ln  \!\left
  (1+\frac   {s}{\mu_D^2} \right ) \Bigg], 
\end{equation}
\begin{equation}
\sigma^{gg}_{q\bar{q}} = \frac{g^4}{54\pi s} \Bigg[ -7 +\frac{3
  \mu_D^2}{s+\mu_D^2} +4 \left (1+\frac {\mu_F^2}{s} \right )
\ln \! \left (1+\frac {s}{\mu_F^2} \right ) \Bigg ]= \frac {32}{9}
\sigma_{gg}^{q\bar{q}}, 
\end{equation}
\begin{equation}
\sigma_{q\bar{q}}^{q'\bar{q}'} = \frac{g^4}{54 \pi (s+\mu_D^2)},
\end{equation}
\begin{equation}
\sigma_{qq'} = \frac{g^4}{36\pi s} \Bigg[
1+\frac{2s}{\mu_D^2}+\frac{\mu_D^2}{s+\mu_D^2} 
- 2 \left (1+\frac {\mu_D^2}{s} \right )  \ln  \! \left
  (1+\frac   {s}{\mu_D^2} \right ) \Bigg].
\label{sigqqp}
\end{equation}

\begin{figure}[htb]
\includegraphics[width=0.7\linewidth]{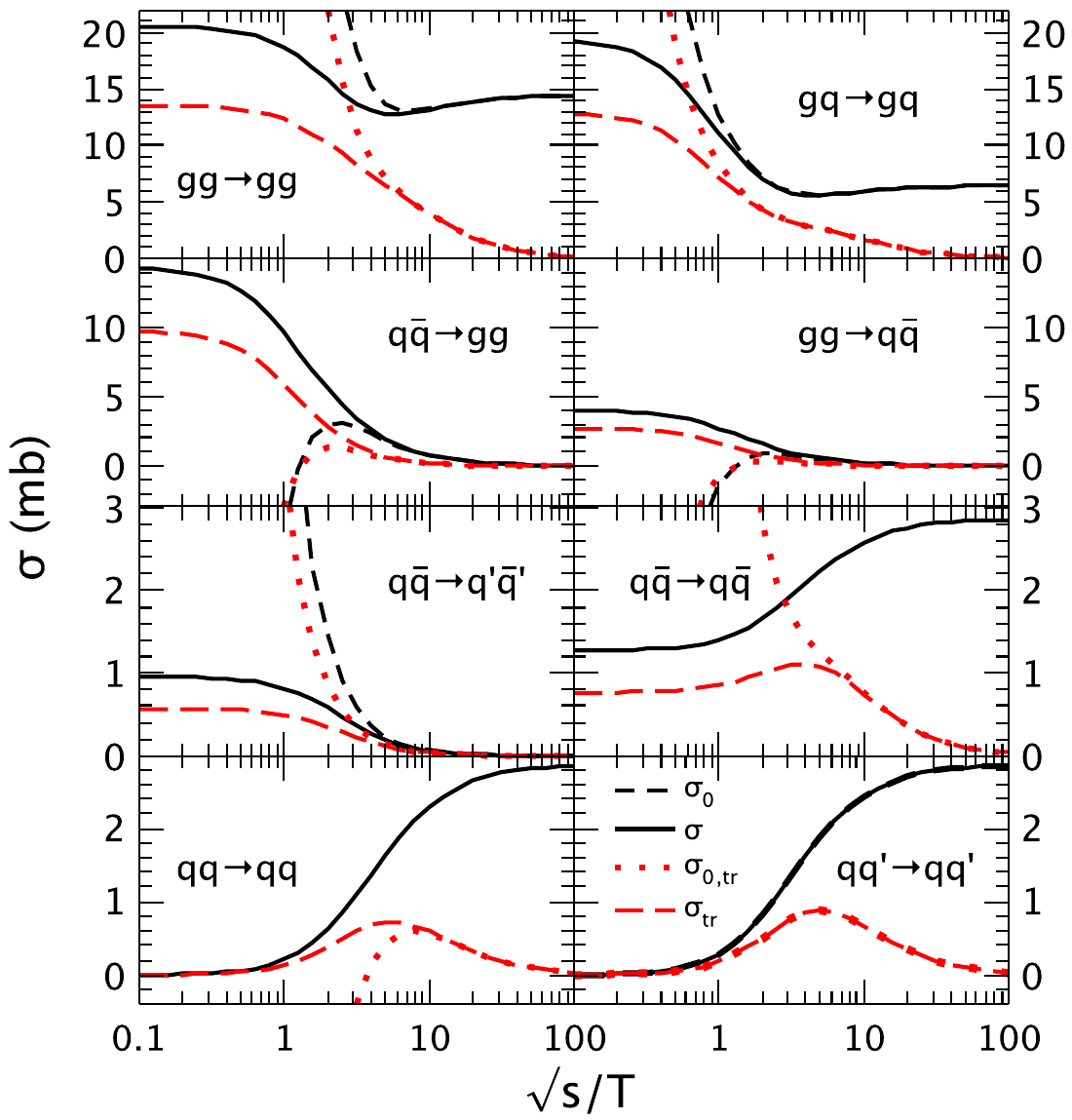}
\caption{Cross section and transport cross section as functions of
  $\sqrt{s}/T$ for each $2\rightarrow   2$ parton scattering channel at
  $T=200\, \mathrm{MeV}$  from the original and modified matrix elements.} 
\label{figSigma}
\end{figure}
   
In Fig.~\ref{figSigma}, we plot the cross section of each $2
\rightarrow 2$ process at $T=200$ MeV and $N_f=3$ as a function of the
scaled two-parton center-of-mass energy $\sqrt s/T$, where 
the total and transport cross sections using the final matrix elements 
($\sigma$ and $\sigma_{\rm tr}$) are compared with those using the
original matrix elements ($\sigma_0$ and $\sigma_{0,\rm tr}$). 
Note that all cross sections here are calculated using the default
parameter values, i.e., $Q=3T$ for Eq.\eqref{alphas} and $\kappa=0.4$
for Eq.\eqref{kappa}. For cross sections obtained from the original
matrix elements,  we see that for $gg \leftrightarrow q\bar{q}$ 
and $qq \rightarrow qq$ channels they are negative at small $s$ and thus
require corrections such as those shown in
Eqs.\eqref{mqq}-\eqref{mqqbarA}. 
In addition, the total cross sections $\sigma_0$ for the $qq\rightarrow
qq$ channel as well as the $q\bar{q}\rightarrow q\bar{q}$ channel are 
infinite and thus cannot be shown in Fig.~\ref{figSigma}, although
their transport cross sections are finite (at non-zero $s$). 
Furthermore, the cross sections $\sigma_0$ of $gg \rightarrow  gg$,
$gq\rightarrow gq$, and $q\bar{q}\rightarrow q'\bar{q}'$ channels 
are all proportional to $1/s$ as $s \rightarrow 0$. 
For total cross sections $\sigma$ obtained from the final matrix
elements with full screenings (solid curves), we see that they are
indeed non-negative and finite at all $s$ values including $s=0$,
despite the appearance of terms such as $1/s$ or even $1/s^3$  
in Eqs.\eqref{siggg}-\eqref{sigqqp}. 
Regarding the transport cross section $\sigma_{\rm tr}$ (long-dashed
curves), we find that  $\sigma_{\rm   tr}/\sigma \rightarrow 0$ as $s
\rightarrow \infty$ for all the channels other than
$q\bar{q}\rightarrow q'\bar{q}'$, meaning that these
finite-temperature $2 \rightarrow 2$ cross sections are very
forwarded-angled at high center-of-mass energies. We also find that
all the  $2 \rightarrow 2$ scatterings become almost isotropic at low
center-of-mass energies ($\sqrt{s}/T  \lesssim 3$), noting that
$\sigma_{\rm tr}=2\sigma/3$ for isotropic cross
sections~\cite{Huovinen:2008te,MacKay:2022uxo}.

\begin{figure}[htb]
\includegraphics[width=0.7\linewidth]{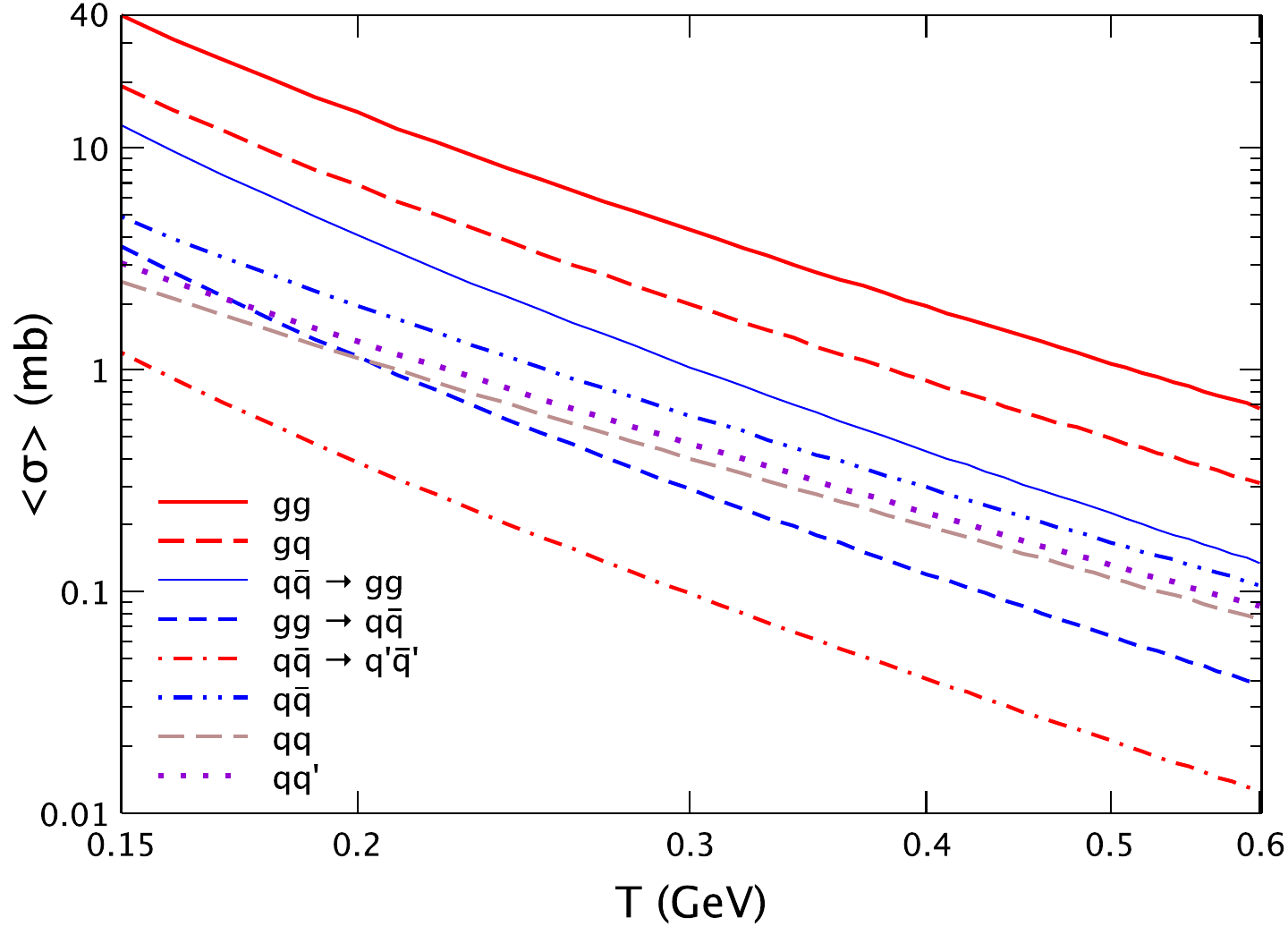}
\caption{Thermal average of the cross section of each
  $2\rightarrow 2$ channel versus the temperature.}
\label{figSigmaThermal}
\end{figure}
 
To see the general magnitudes of the cross sections, we plot in
Fig.~\ref{figSigmaThermal} the thermally averaged cross
section, defined as~\cite{Kolb:1983sj,MacKay:2022uxo}
\begin{equation}
\langle \sigma \rangle = \frac{1}{32} \int_0^{\infty}
\diff v\, \left [ v^4 K_1(v) + 2v^3\, K_2(v) \right ] \sigma(v).
\label{sigmaThermal}
\end{equation}
We see big differences among the channels at a given
temperature, with the elastic $\langle \sigma_{gg}\rangle$ being the
largest and  the inelastic $\langle \sigma_{q\bar{q}}^{q'\bar{q}'}
\rangle$ being the smallest. 
For the other two inelastic channels, 
$\langle \sigma_{q\bar{q}}^{gg} \rangle$ is the 3rd largest among all
the thermally averaged cross sections, while
$\langle\sigma_{gg}^{q\bar   q} \rangle =9 \langle \sigma_{q\bar
  q}^{gg} \rangle/32$ is much smaller. We also see that all of the
average cross sections strongly decrease with the temperature,
numerically roughly  as $1/T^3$ within the shown temperature range in
Fig.~\ref{figSigmaThermal}. At the QCD phase transition temperature
$T_c \simeq 156$ MeV, $\langle  \sigma_{gg}\rangle$ can be as large as
$\sim 34$ mb, while $\langle \sigma _{q\bar   q}^{q^{\prime} \bar
  q^{\prime}} \rangle$ is only $\sim 1.0$ mb.

\section{Results on shear viscosity}
\label{Results}

\begin{figure}[htb]
\includegraphics[width=0.8\linewidth]{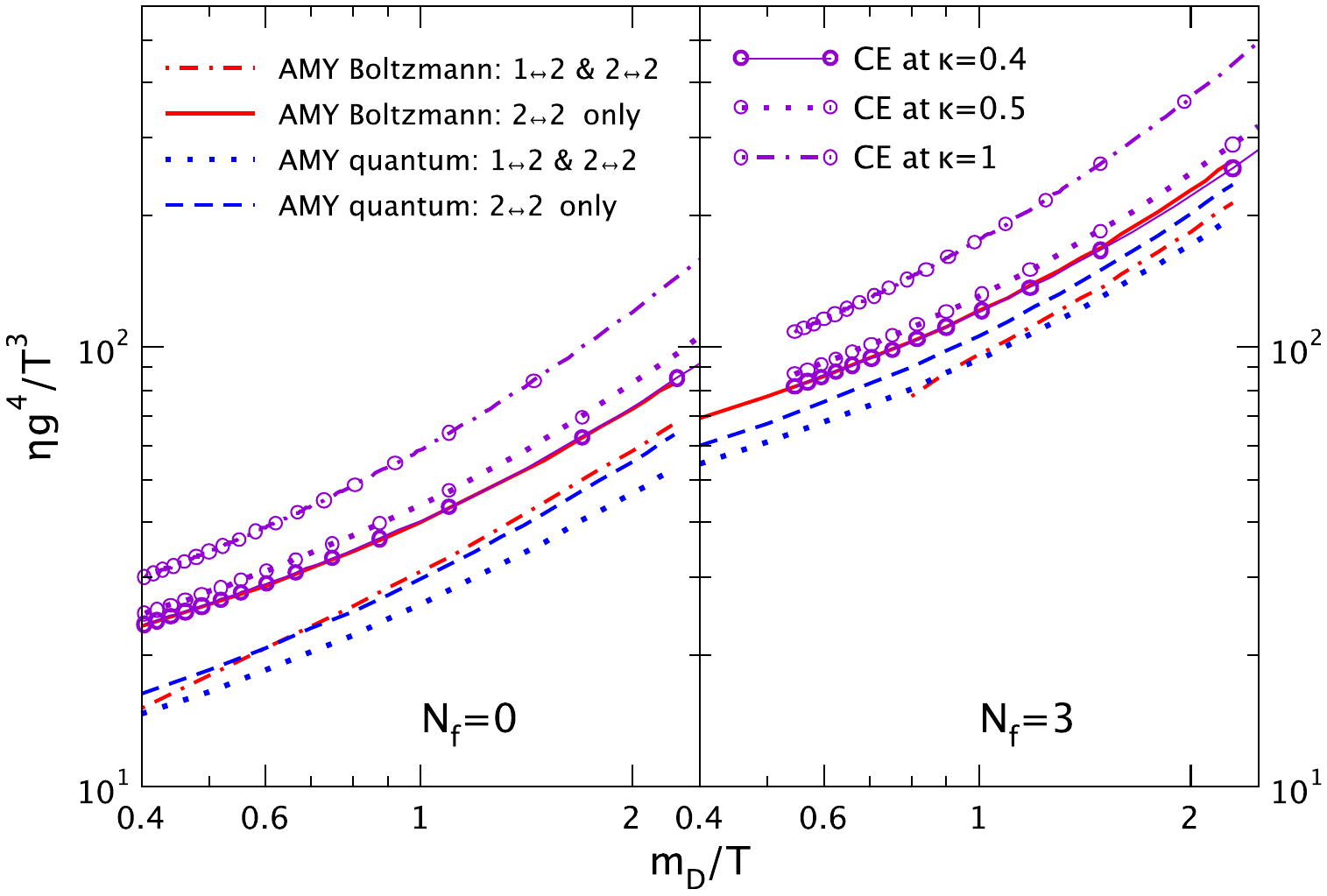}
\caption{Chapman-Enskog results of $\eta \, g^4/T^3$ at different
  $\kappa$ values (curves with circles) as functions of $m_D/T$ for
  $N_f=0$ (left panel) and $3$ (right panel) compared with the
  leading-order AMY results for quantum statistics or Boltzmann
  statistics with or without $1\leftrightarrow 2$
  processes~\cite{Moore:2025}.}
\label{vsAMY}
\end{figure}

Applying the CE analytical expression of Eq.\eqref{etaNf} to the
$2\leftrightarrow 2$ scattering cross sections using the final matrix
elements of Eqs.\eqref{mqqp0}-\eqref{mqqbarB}, we calculate the
corresponding shear viscosity $\eta$ of the quark-gluon plasma at
finite temperature.   Figure~\ref{vsAMY} shows our CE results of $\eta
\,g^4/T^3$ (curves with circles) as functions of $m_D/T$ for $N_f=0$ and
$3$; the leading-order results from the AMY 
framework~\cite{Arnold:2000dr, Arnold:2003zc} for quantum statistics
or Boltzmann statistics with or without $1\leftrightarrow 2$
processes~\cite{Moore:2025} are also shown for comparison. As
expected, the AMY result of $\eta \, g^4/T^3$ decreases after including
$1\leftrightarrow 2$
processes~\cite{Arnold:2000dr,Arnold:2003zc,Teaney:2009qa}. We also
see that the AMY results for Boltzmann statistics are mostly 
higher than those for quantum statistics. 

Since our CE calculations
include $2\leftrightarrow 2$ processes with Boltzmann statistics, it
is the most proper to compare our CE results with the AMY results for
Boltzmann statistics and $2\leftrightarrow 2$ processes (solid curves
without symbols). For $\kappa=1$ (i.e., screening the  matrix elements
with thermal masses $m_D^2$ or $m_F^2$), we see that the CE results of
shear viscosity are significantly higher than the corresponding AMY
results;  this is consistent with the earlier
finding~\cite{Arnold:2003zc} that using thermal masses for  the
screening overestimates the quark diffusion constant by $\sim 
50\%$ compared to the full AMY result that uses momentum-dependent
self-energies for the screening.    Therefore, we modify the screening
by adding the $\kappa$ parameter,  i.e., screening the matrix elements
with thermal masses $\mu_D^2=\kappa m_D^2$ or $\mu_F^2=\kappa m_F^2$
as shown in Eq.\eqref{kappa}. Note that a similar $\kappa$
factor has been used in an earlier work~\cite{Gossiaux:2008jv} so that
the pQCD Born calculation with $t$ replaced with $t-\kappa m_D^2$
(with $\kappa \sim 0.2$ there) for the gluon propagator gives the
same energy loss as a more complete method. In Fig.~\ref{vsAMY}, we
see that $\kappa=0.4$ allows the CE results of $\eta\, g^4/T^3$ to
agree well with the AMY results for Boltzmann statistics and
$2\leftrightarrow 2$ processes (solid lines) for both the $N_f=0$ and
$N_f=3$ cases. This is why we use  $\kappa=0.4$ by default throughout
this study. 

\begin{figure}[htb]
\includegraphics[width=0.7\linewidth]{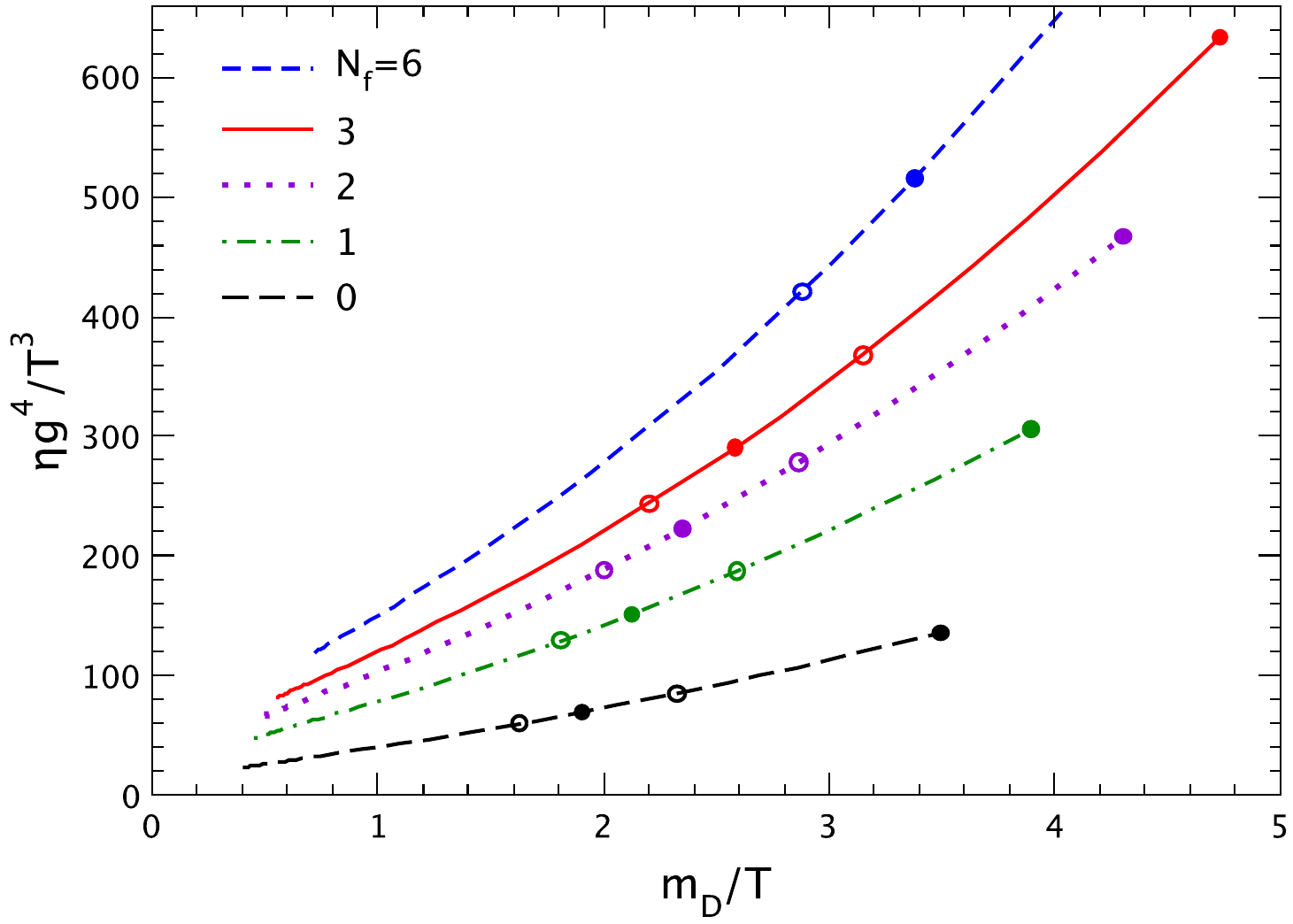}
\caption{The CE results of $\eta \, g^4/T^3$ at $\kappa=0.4$ for
  different $N_f$ values as functions of $m_D/T$; filled circles
  represent the default results (taking momentum transfer $Q=3T$) at
  $T=150$ and $600$ MeV, while open circles represent the results
  taking $Q=2\pi T$.} 
\label{vsMD}
\end{figure}

We show in Fig.~\ref{vsMD} the CE results of $\eta \, g^4/T^3$ at
different $N_f$ values as functions of $m_D/T$, where we take
$\kappa=0.4$. The shear viscosity has been presented this way in the
earlier AMY works~\cite{Arnold:2000dr,Arnold:2003zc}, since $\eta \,
g^4/T^3$ only depends on the value of $m_D/T$ (that is proportional to
the strong coupling $g$) but not on the choice of momentum scale $Q$
used in $g$.  Also note that the $m_D/T$ range in Fig.~\ref{vsMD}
covers much higher values because we present the CE results for
temperatures down to 150 MeV (starting from $10^{18}$ GeV). 
To illustrate these features, in Fig.~\ref{vsMD} we use the filled
circles to represent the CE results at $T=150$ MeV and $600$ MeV 
from taking the momentum transfer as $Q=3T$, while we use open
circles to represent the CE results from taking $Q=2\pi
T$~\cite{Blaizot:2001tn}, where the $T=150$ MeV point is located at a 
higher $m_D/T$ value than the corresponding $T=600$ MeV point. Indeed
we see that the CE results for the two choices of $Q$ lie on the same
curve. The $\eta \, g^4/T^3$ values are seen to increase with $N_f$
(at any given value of $m_D/T$), similar to the AMY
results~\cite{Arnold:2003zc}. 

\begin{figure}[htb]
\includegraphics[width=0.7\linewidth]{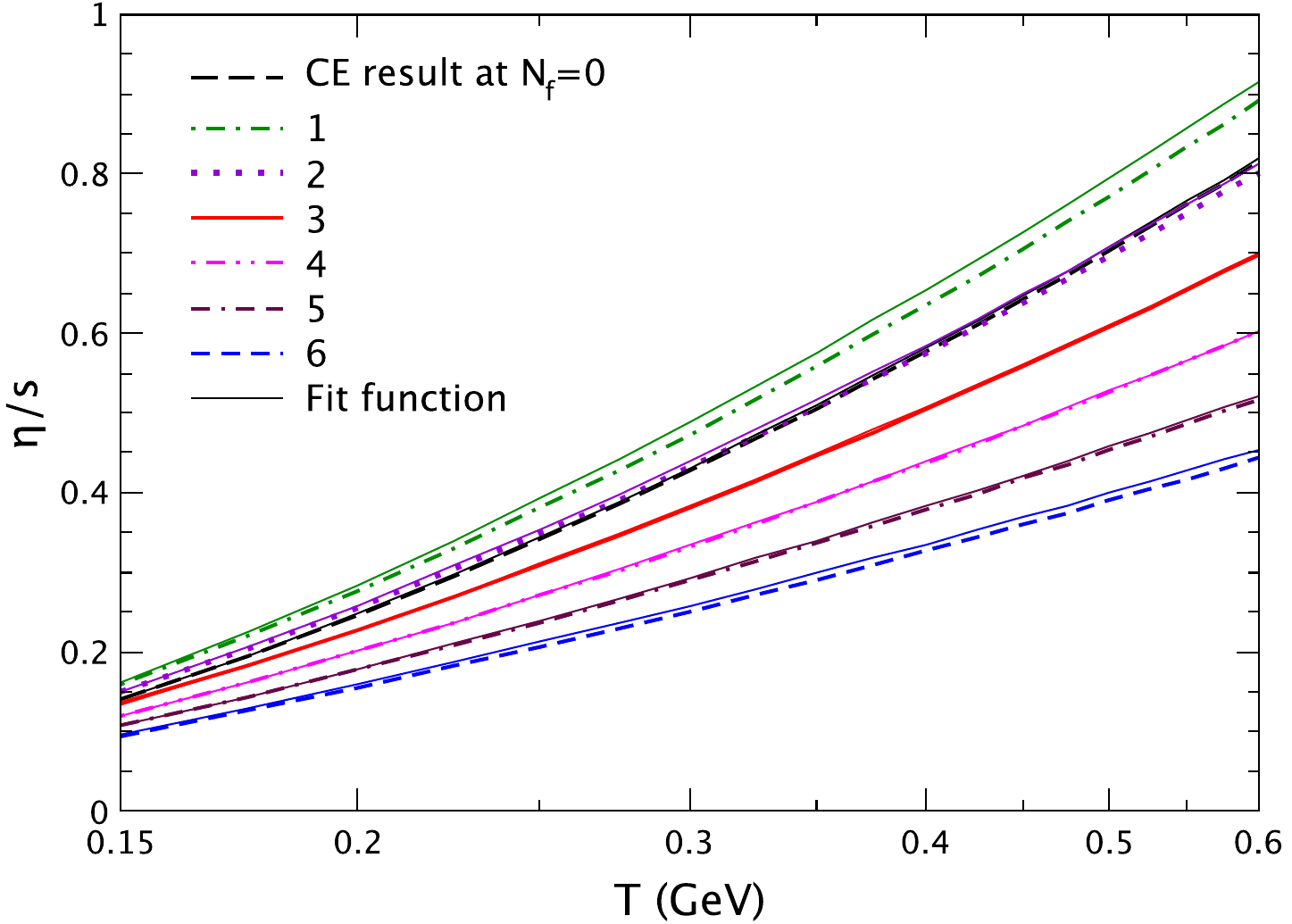}
\caption{Shear viscosity-to-entropy density ratio as a function of
  temperature for a quark gluon plasma at different $N_f$
  values; fit functions over the full $m_D/T$ range are shown as thin
  solid curves.} 
\label{etaovers}
\end{figure}

Figure~\ref{etaovers} shows our CE results of the shear
viscosity-to-entropy density ratio for different numbers of quark
flavors for the default choice of $Q=3T$ and $\kappa=0.4$. 
We see that the $\eta/s$ ratio increases strongly with 
temperature, and its lowest value (at the critical temperature $T \simeq
156$ MeV) is quite low at $\eta/s \sim 0.15$ for $N_f=3$, less than
twice the conjectured lower bound of $1/(4\pi)$~\cite{Kovtun:2004de}. 
Regarding the $N_f$ dependence, we see that  the $\eta/s$ ratio first 
increases when $N_f$ changes from 0 to 1, then it decreases with
$N_f$. This is understandable because the average $gg$ elastic cross
section is the biggest among all the $2 \rightarrow 2$ scattering 
channels, which leads to a lower $\eta/s$ value for pure gluon gas than
that for $N_f=1$. On the other hand, as $N_f$ further increases the
number of individual channels such as $gg \leftrightarrow q_i \bar
q_i$ and $q_i q_j \rightarrow q_i q_j$ increases while $\alpha_s$
also becomes bigger; they increase the scatterings rate per parton and
thus decrease the $\eta/s$.  
In this study we use the entropy density for a massless QGP
with Boltzmann statistics, which is given by $s =16(4+3N_f)T^3/\pi^2$;
note that the Boltzmann and quantum entropy densities differ by $<8\%$
for $N_f \in [0, 6]$.  

It would be useful to have an explicit functional form for the CE
shear viscosity results shown in Figs.~\ref{vsMD}-\ref{etaovers}.  
For the simplest $N_f=0$ case (i.e., pure gluon gas), the shear
viscosity is given by the single-species
result~\cite{Plumari:2012ep,MacKay:2022uxo}:
\begin{equation}
\eta =\frac{61440\, T}{\int_0^{\infty} \diff v\, v^6 \left[(3v^2 +
    4)K_3(v) - 6v\, K_2(v)\right] \sigma_{\rm tr}^{gg \rightarrow gg}(v)}. 
\label{singleSpecies}
\end{equation}
For two-gluon elastic scatterings, which matrix element at  finite
temperature is given by Eq.\eqref{mgg0} or Eq.\eqref{mgg}, the 
above integral in Eq.\eqref{singleSpecies} does not yield an explicit
function. We thus fit all the $\eta \,g^4/T^3$ results in
Fig.~\ref{vsMD} as the following function of 
$m_D/T$ and $N_f$: 
\begin{equation}
\left ( \frac{\eta \, g^4}{T^3} \right )_{\rm fit}
=17.0\; (1 + 2.25 N_f)^{0.623} + \left ( 60.5 - 37.7 e^{-0.754 N_f}
\right ) \left ( \frac{m_D}{T} \right )^{-3.76 + 1.55 \ln \small (N_f +
  26.4 \small )}. 
\label{wideFit}
\end{equation}
One can then obtain the corresponding fit function for
$\eta/s(T, N_f)$, which are shown as the thin solid lines in
Fig.~\ref{etaovers}. For example, the fit function $(\eta/s)_{\rm
  fit}=[2.89 + 2.68 (m_D/T)^{1.48}]/g^4$ for $N_f=3$ while given by 
$[2.62+3.52(m_D/T)^{1.31}]/g^4$ for $N_f=0$, where we note that 
$g$ is a function of both $T$ and $N_f$.  We see in
Fig.~\ref{etaovers}  that the fit functions describe the exact CE
results quite well for each $N_f$ value within $[0, 6]$. 
When we just fit the $\eta \, g^4/T^3$ values in Fig.~\ref{vsMD}
within a narrow range $T\in [0.15,0.6]$ GeV, instead of the full range
of $m_D/T$ that corresponds to $T \in [0.15,10^{18}]$ GeV, we get the 
following fit function that almost perfectly matches the exact CE
$\eta/s$ results in Fig.~\ref{etaovers}:
\begin{equation}
\left ( \frac{\eta \, g^4}{T^3} \right )_{\rm narrow-fit}
=24.7\; (1 + 2.63 N_f)^{0.596} + \left ( 45.5 - 28.3 e^{-0.659 N_f}
\right ) \left ( \frac{m_D}{T} \right )^{0.808+0.384\ln \small (N_f +
  5.91 \small )}. 
\label{narrowFit}
\end{equation}
For example, this narrow-fit function gives the $\eta/s$ ratio as
$(\eta/s)_{\rm   narrow-fit}=[4.31 + 1.97 (m_D/T)^{1.65}]/g^4$ for
$N_f=3$ and   
$[3.81+2.65(m_D/T)^{1.49}]/g^4$ for $N_f=0$. 

\begin{figure}[htb]
\includegraphics[width=0.6\linewidth]{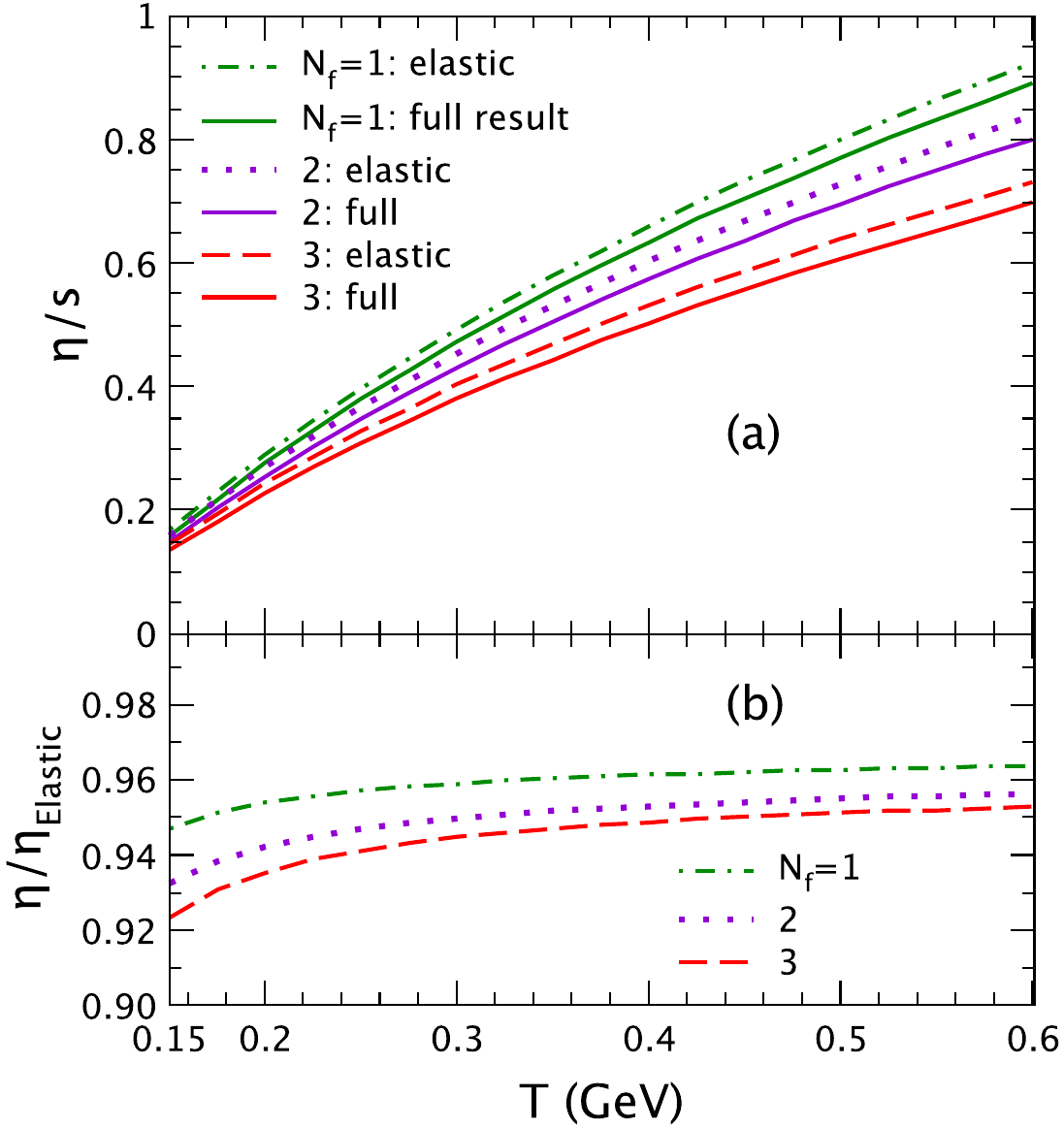}
\caption{(a) The $\eta/s$ ratio of QGP at different $N_f$,
  where the full result includes both elastic and inelastic
  $2\leftrightarrow 2$ scatterings. (b) Ratio of the full result
  of shear viscosity to the shear viscosity from only elastic scatterings.}  
\label{etas-elastic}
\end{figure}

Next we examine the effects of inelastic $2 \rightarrow 2$ scatterings
on the shear viscosity. Figure~\ref{etas-elastic}(a) shows the $\eta/s$
ratio of a massless QGP with (solid lines, named as full result) or
without inelastic $2 \rightarrow 2$ scatterings for $N_f=1,2,3$. We
see that the inelastic scatterings of $gg \leftrightarrow q \bar q$
and $q \bar q \rightarrow  q^\prime \bar q^\prime$ lead to a modest
decrease of the shear viscosity. The ratios of the full shear viscosity
to the shear viscosity from elastic scatterings only are shown
in  Fig.~\ref{etas-elastic}(b), and we see that the decrease of shear 
viscosity due to inelastic scatterings range from 4\% to about 8\% for 
$N_f=1,2,3$. We also see that the effect from inelastic scatterings
increases with $N_f$, partly due to the increasing number of
individual channels as mentioned earlier.

\section{Discussions}
\label{Discussions}

\begin{figure}[htb]
\includegraphics[width=0.6\linewidth]{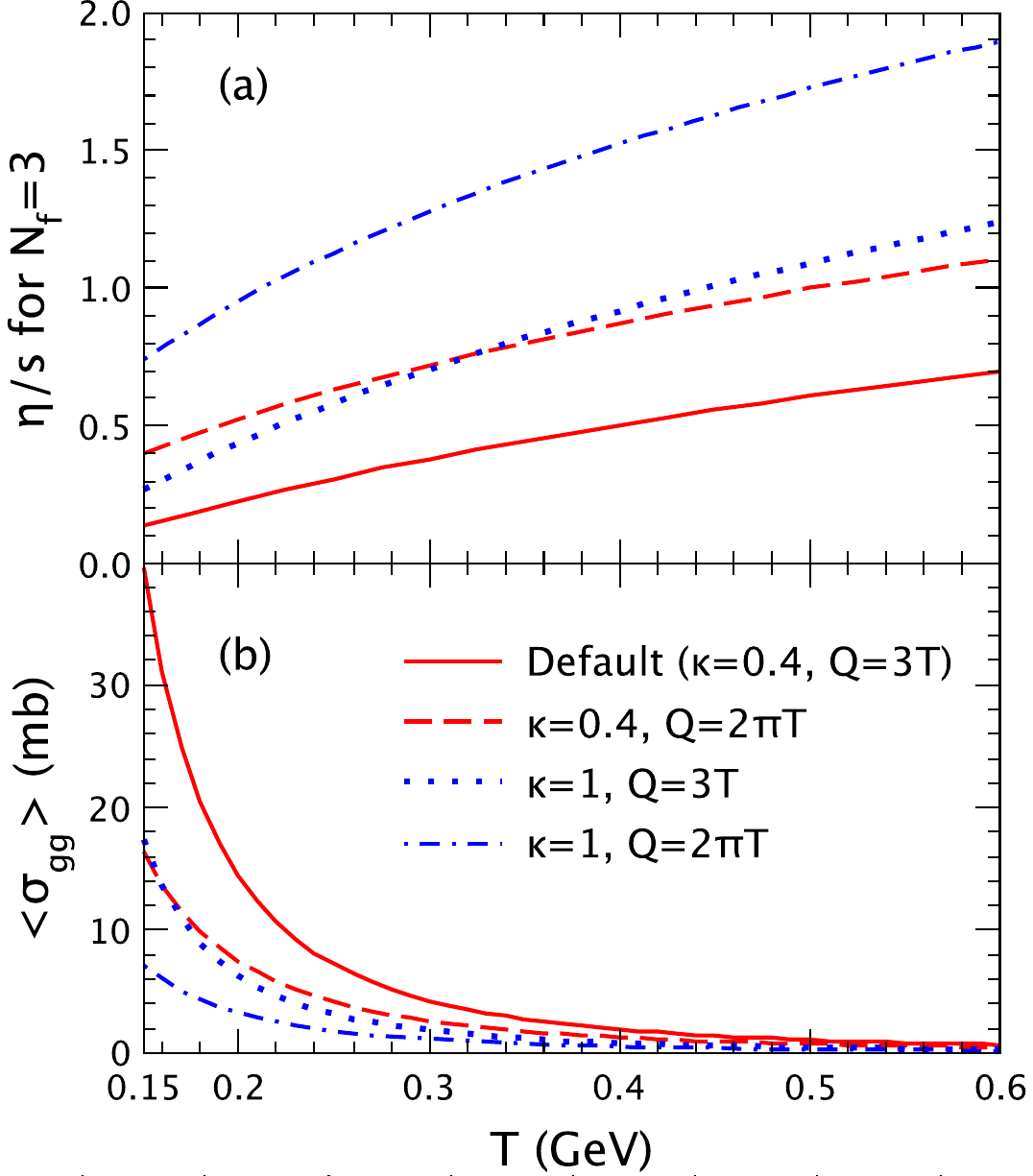}
\caption{(a) The $\eta/s$ ratio and (b) the
  thermally averaged cross section for $gg$ elastic scatterings versus
  temperature for $N_f=3$ QGP for the choice of momentum transfer
  $Q=3T$ or $Q=2\pi T$ and the screening coefficient $\kappa=0.4$ or
  $1$.}
\label{etas-vs-q}
\end{figure}
  
We have introduced a screening coefficient $\kappa$ in 
Eq.\eqref{kappa} and then taken $\kappa=0.4$ as shown in
Fig.~\ref{vsAMY}; a main purpose is to take into account the
difference between our simpler approach that uses thermal masses to
screen the matrix elements  and the AMY result that uses
self-energies for the screening.  Another parameter is the momentum
transfer $Q$ that is used in the strong coupling constant of 
Eq.\eqref{alphas}. We demonstrate in Fig.~\ref{etas-vs-q}(a) the
effect of the choice of momentum transfer $Q$ as well as the effect
of screening coefficient $\kappa$ on the $\eta/s$ ratio for three
quark flavors, by comparing our default choice $Q/T=3$ with
$Q/T=2\pi$~\cite{Blaizot:2001tn} and comparing the default choice
$\kappa=0.4$ with $\kappa=1$. Although the choice of $Q$ does not
affect $\eta \, g^4/T^3$ as a function of $m_D/T$ (as shown in
Fig.~\ref{vsMD}), it has a large effect on
$\eta/s$~\cite{Ghiglieri:2018dib}, where a  smaller $Q/T$ value leads
to a lower $\eta/s$ at the same temperature. This is mainly due to the
fact that a lower $Q$ leads to  a bigger $g$ and thus a bigger cross
section (due to the overall $g^4$ factor in each cross section); this
is shown in Fig.~\ref{etas-vs-q}(b)  for the $gg$ elastic scattering
channel as an example. This $Q$-dependence can also been seen from the
explicit fit functions $(\eta/s)_{\rm   fit}$ or $(\eta/s)_{\rm
  narrow-fit}$ given earlier, where $\eta/s \propto 1/g^4$ is the
leading dependence on $g$,  although there is also a sub-leading
dependence on $g$ as $m_D/T$ to the power of less than $2$. Regarding
the $\kappa$ dependence, we see in Fig.~\ref{etas-vs-q}(a) that a
smaller $\kappa$ leads to a lower  $\eta/s$, similar to the effect of
$Q/T$. This is because a small $\kappa$ leads to a bigger scattering
cross section $\sigma$, as shown in Fig.~\ref{etas-vs-q}(b), and thus
a lower viscosity, noting that $\eta \propto T/\sigma$ for isotropic
scatterings.

\begin{figure}[htb]
\includegraphics[width=0.7\linewidth]{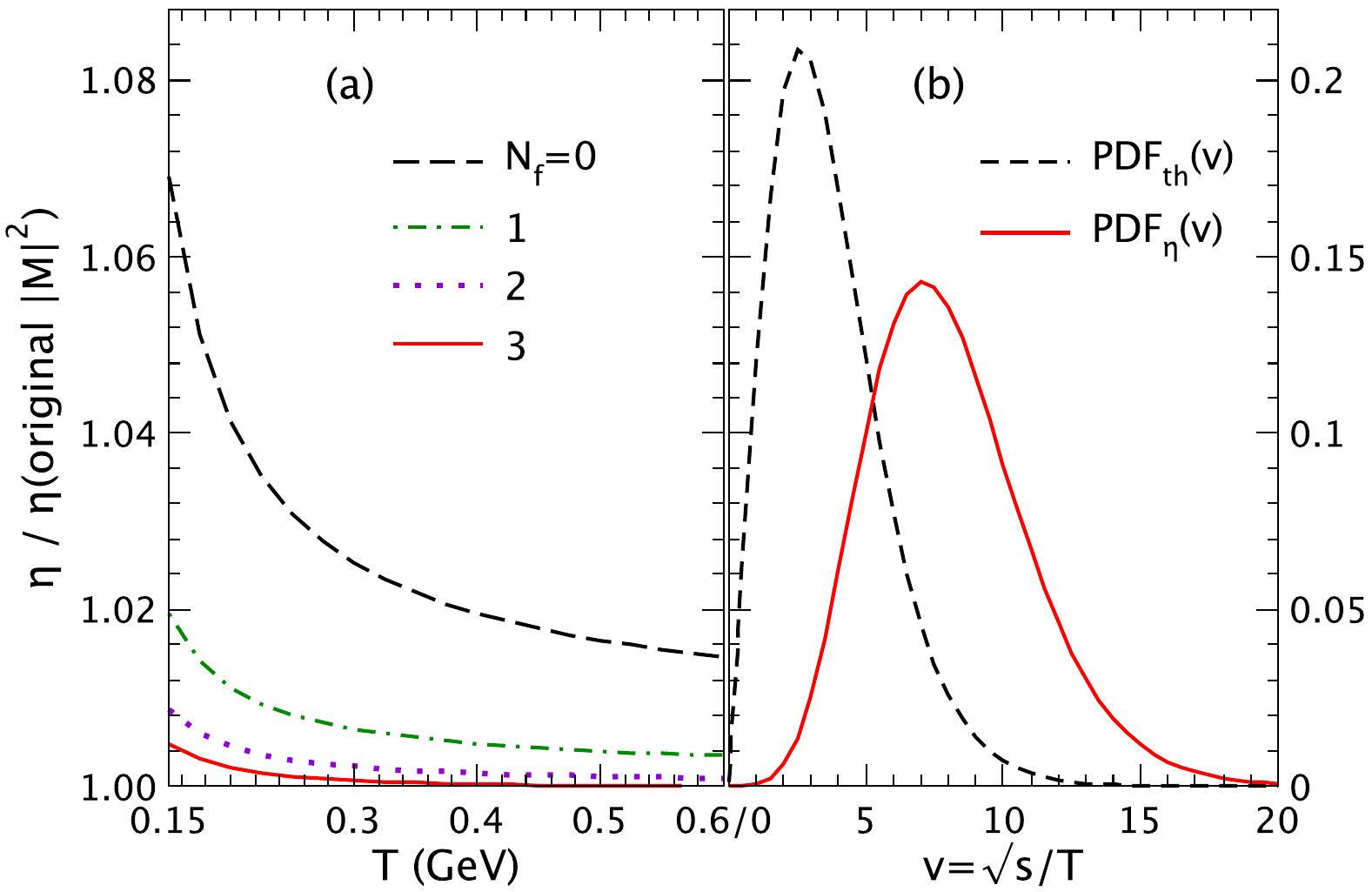} 
\caption{(a) Ratio of the shear viscosity with fully screened matrix 
  elements over that with the original matrix elements versus
  temperature at different $N_f$. (b) The PDF of $v$ for calculating
  the thermal average and for calculating the shear viscosity of a
  single massless particle species.} 
\label{screening}
\end{figure}

In our method of using scaled thermal masses to screen the matrix
elements and cross sections, we have added extra screenings to the
original matrix elements in Eqs.\eqref{mgg0}-\eqref{mqqbar0B}. The
extra screenings in Eqs.\eqref{mqq}-\eqref{mqqbarA} are necessary to
remove the unphysical negative values of the original matrix elements. 
The other extra screenings in Eqs.\eqref{mqqb}-\eqref{mqqbarB} make
the total cross section of the individual channels finite at any
two-parton collision energy. Finite cross sections are necessary for 
parton cascade models
\cite{Zhang:1997ej,Lin:2004en,Xu:2007ns,Cassing:2009vt,Plumari:2011mk} 
and the QCD effective kinetic theory approach~\cite{Kurkela:2018vqr};
however, they are not necessary for 
the calculation and finiteness of shear viscosity in the
Chapman-Enskog method or the AMY framework. 
Figure~\ref{figSigma} also shows that the changes to the total and
transport cross sections due to the extra screenings  are often large, 
especially at low two-parton energies $\sqrt{s}/T$. Therefore, we
investigate in Fig.~\ref{screening} the effect of the extra screenings
on the shear viscosity. Figure~\ref{screening}(a) shows the ratio of
the shear viscosity with extra screenings to that without, and we 
see that the extra screenings only lead to a small change of the shear 
viscosity, e.g., by less than 1\% for $N_f=3$. 

To understand why the effect is so small, we plot the following
probability density function used in calculating the thermal average
in Eq.\eqref{sigmaThermal}: 
\begin{equation}
{\rm PDF}_{\rm th}(v)=\left [ v^4 K_1(v)+ 2v^3\, K_2(v) \right ]/32, 
\end{equation} 
shown as the dashed line in Fig.~\ref{screening}(b). 
It shows that the center-of-mass energy of two partons peaks at the
$v$  value of several $T$ with a finite width so that there are few
scatterings at very low $v$ or $\sqrt{s}/T$ values. 
The solid curve in Fig.~\ref{screening}(b) shows the probability
density function used in calculating the CE shear viscosity of a
single massless particle species via Eq.\eqref{singleSpecies}:
\begin{equation}
{\rm PDF}_{\rm \eta}(v)= v^6 \left[ (3v^2 +     4)K_3(v) - 6v\,
  K_2(v)\right]/76800.  
\end{equation}
Note that this function also appears in certain Chapman-Enskog matrix
elements such as the $\Tilde{c}_{0}[i]$ term given by Eq.\eqref{c0g}. 
We see in Fig.~\ref{screening}(b) that the ${\rm PDF}_{\rm
  \eta}(v)$ function has its peak at a much higher $v$ value and its low-$v$
part much suppressed than the thermal average PDF. As a result, the
large changes of cross sections at low $v$ values due to our extra
screenings lead to only a small effect on the shear viscosity of the
QGP. This reflects the fact that the shear viscosity is related to
finite momentum transfer and thus parton collisions at low
center-of-mass energies do not contribute much.

We also note that our results in this study includes all the
$2\leftrightarrow 2$ interactions. However, it does not include
$1\leftrightarrow 2$ interactions, which would decrease the
shear viscosity as shown by the AMY results in Fig.~\ref{vsAMY}. 
In addition, we use Boltzmann statistics, while the quantum statistics
would also lower the shear viscosity. Furthermore, next-to-leading order
corrections~\cite{Ghiglieri:2018dib} to the leading-order AMY results
have been found to significantly decrease the $\eta/s$ value at finite
temperatures. This study demonstrates the mapping between the QGP
shear viscosity from the Chapman-Enskog method and specific parton
two-body cross sections.  However, parton scattering cross sections at
finite temperature suffer from large uncertainties. Further
comparisons with results from the AMY framework could shed light on
how various scattering channels behave at finite temperatures and thus
lead to better ways to model those cross sections.

\section{Conclusion}
\label{Conclusion}

We use the Chapman-Enskog method to study the shear viscosity of 
quark-gluon plasma at finite temperature and its relation to parton
$2\leftrightarrow 2$ cross sections. In a recent work, we have derived
the analytical expression for the shear viscosity of a massless
quark-gluon plasma with Boltzmann statistics under all two-body
elastic and inelastic parton scatterings, where the differential cross
section of each scattering channel can be arbitrary. Here we 
apply this general expression to parton cross sections
at finite temperature, which are based on perturbative-QCD and
screened with scaled thermal masses  $\sqrt{\kappa} m_D$ and
$\sqrt{\kappa} m_F$. At $\kappa=1$, our Chapman-Enskog results on
$\eta \, g^4/T^3$ as functions of $m_D/T$ are qualitatively similar
to but quantitatively higher than the corresponding leading-order
results from the AMY framework, i.e., the AMY results for Boltzmann
statistics under  $2\leftrightarrow 2$  scatterings. 
On the other hand, our Chapman-Enskog results at $\kappa=0.4$ 
match well the corresponding AMY results, partly because it takes care
of the difference between using thermal masses (our method) and
using self-energies (the AMY framework) to screen the cross sections. 

We have added extra screenings to regulate parton cross
sections so that they are non-negative and finite, and these extra
screenings have a large effect on parton cross sections at low
two-parton center-of-mass energies or low $v$ (defined as
$\sqrt{s}/T$). However, they have a very small effect on the shear 
viscosity, because the shear viscosity depends on weighting functions
${\rm PDF}(v)$ that peak at finite $v$ while being significantly
suppressed  at low $v$ values. We also show that the shear
viscosity-to-entropy density ratio $\eta/s$ is very sensitive to the
choice of the momentum scale $Q$ (in terms of $T$) used in the strong
coupling constant, although the choice of $Q$ does not affect $\eta \,
g^4/T^3$ as a function of  $m_D/T$.  We find that the choice of
$Q=\sqrt{\langle -t \rangle}=3T$ leads to a low shear viscosity
$\eta/s \sim 0.15$ for $N_f=0$ or 3 at the QCD phase transition
temperature $T_c \simeq 156$ MeV.  
In addition, we provide explicit fit functions for our results of $\eta
\, g^4/T^3$ and $\eta/s$ for the default parameter values  (i.e.,
$\kappa=0.4$ and $Q=3T$). This study provides a
demonstration of relating the QGP shear viscosity at finite
temperature to adjustable parton scattering cross sections, which is
expected to be useful for studies of the quark-gluon matter from
kinetic theory-based approaches.

\section*{Acknowledgments}
We thank P. Arnold, G. Moore and M.A. Ross for helpful
discussions. This work has been supported by the National Science  
Foundation under Grant No. 2310021.

\bibliography{refs}

\end{document}